\documentclass[journal=ancac3,manuscript=article,layout=twocolumn]{achemso}
\usepackage{amsmath,amssymb,amsfonts}
\usepackage{newpxtext}
\usepackage{newpxmath}
\usepackage[utf8]{inputenx}
\usepackage{graphicx}
\usepackage[dvipsnames]{xcolor}
\usepackage[pdftex]{hyperref}

\DeclareMathAlphabet{\mathcal}{OMS}{cmsy}{m}{n}

\hypersetup{
	pdfauthor={Gabriele Calabrese},
	bookmarksnumbered=true,
	pdftitle={Radius-Dependent Homogeneous Strain in Uncoalesced GaN Nanowires},
	colorlinks,
	citecolor=blue,
	linkcolor=blue,
	urlcolor=blue}

\usepackage{xspace}
\usepackage{soul}

\DeclareUnicodeCharacter{2212}{-}

\title{Radius-Dependent Homogeneous Strain in Uncoalesced GaN Nanowires}

\author{G.~Calabrese}
\affiliation{Paul-Drude-Institut für Festkörperelektronik, Leibniz-Institut im Forschungsverbund Berlin~e.\,V., Hausvogteiplatz 5--7, 10117 Berlin, Germany}
\alsoaffiliation{Dipartimento di Ingegneria dell’Informazione,
Università di Pisa, Pisa 56122, Italy}
\email{calabrese@pdi-berlin.de}
\author{D.~van Treeck}
\affiliation{Paul-Drude-Institut für Festkörperelektronik, Leibniz-Institut im Forschungsverbund Berlin~e.\,V., Hausvogteiplatz 5--7, 10117 Berlin, Germany}
\author{V.~M.~Kaganer}
\affiliation{Paul-Drude-Institut für Festkörperelektronik, Leibniz-Institut im Forschungsverbund Berlin~e.\,V., Hausvogteiplatz 5--7, 10117 Berlin, Germany}
\author{O.~Konovalov}
\affiliation{European Synchrotron Radiation Facility, 71 avenue des Martyrs, 38043 Grenoble, France}
\author{P.~Corfdir}
\affiliation{Paul-Drude-Institut für Festkörperelektronik, Leibniz-Institut im Forschungsverbund Berlin~e.\,V., Hausvogteiplatz 5--7, 10117 Berlin, Germany}
\alsoaffiliation{Present address: ABB Corporate Research, 5405 Baden-Dättwil,
Switzerland}
\author{C.~Sinito}
\affiliation{Paul-Drude-Institut für Festkörperelektronik, Leibniz-Institut im Forschungsverbund Berlin~e.\,V., Hausvogteiplatz 5--7, 10117 Berlin, Germany}
\alsoaffiliation{Present address: Attolight AG, EPFL Innovation Park, Bât.\ D, 1015 Lausanne, Switzerland}
\author{L.~Geelhaar}
\author{O.~Brandt}
\author{S.~Fernández-Garrido}
\affiliation{Paul-Drude-Institut für Festkörperelektronik, Leibniz-Institut im Forschungsverbund Berlin~e.\,V., Hausvogteiplatz 5--7, 10117 Berlin, Germany}
\alsoaffiliation{Grupo de electrónica y semiconductores, Dpto.\ Física Aplicada, Universidad Autónoma de Madrid, C/ Francisco Tomás y Valiente 7, 28049 Madrid, Spain}
\email{sergio.fernandezg@uam.es}

\begin{document}

\begin{tocentry}
\vspace*{-0.14cm}
\centerline{\includegraphics[width=0.9\textwidth]{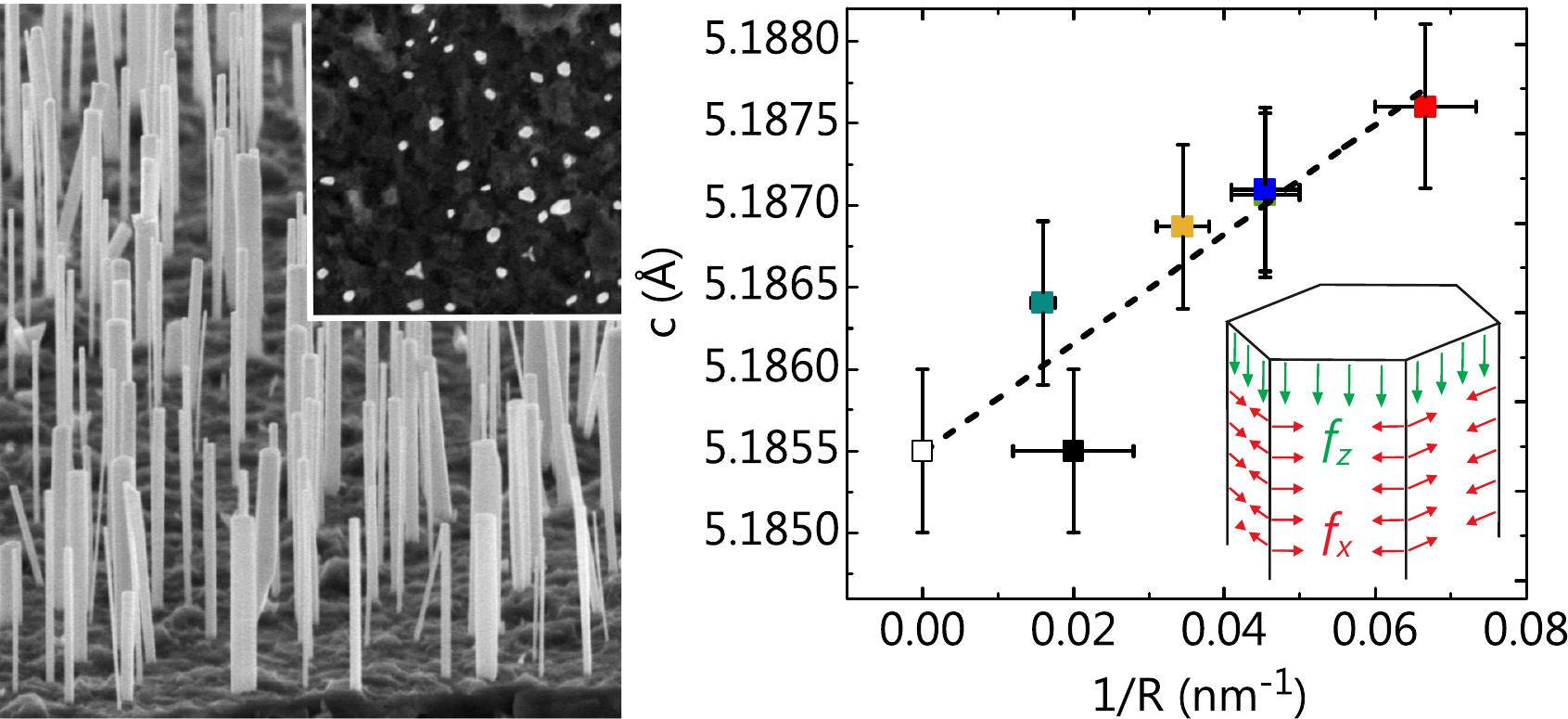}}
\end{tocentry}

\begin{abstract}
{\small We investigate the strain state of ensembles of thin and nearly coalescence-free self-assembled GaN nanowires prepared by plasma-assisted molecular beam epitaxy on Ti/Al$_{2}$O$_{3}(0001)$ substrates. The shifts of Bragg peaks in high-resolution X-ray diffraction profiles reveal the presence of a homogeneous tensile strain in the out-of-plane direction. This strain is inversely proportional to the average nanowire radius and results from the surface stress acting on the nanowire sidewalls. The superposition of strain from nanowires with different radii in the same ensemble results in a broadening of the Bragg peaks that mimics an inhomogeneous strain on a macroscopic scale. The nanowire ensembles show a small blueshift of the bound-exciton transitions in photoluminescence spectra, reflecting the existence of a compensating in-plane compressive strain, as further supported by grazing incidence x-ray diffraction measurements carried out at a synchrotron. By combining X-ray diffraction and photoluminescence spectroscopy, the surface stress components $f_{x}$ and $f_{z}$ of the air-exposed GaN$\{1\bar100\}$ planes that constitute the nanowire sidewalls are determined experimentally to be 2.25 and $-0.7$~N/m, respectively.}

\end{abstract}
\maketitle

\section{Introduction}

Due to the exceptionally large surface-to-volume ratio of nanowires (NWs), surface effects play a key role for their properties. For instance, depending on the material system, nonradiative recombination at the NW surfaces might become a critical issue.\citep{Schlager2008} Hence, proper surface passivation is typically required for enhancing the performance of NW-based devices.\citep{Zhao2015, Varadhan2017} In addition, surface states induce Fermi level pinning at the NW sidewalls, which causes the bending of electronic bands. The consequent formation of a space charge region controls both the electrical transport and optical properties of thin NWs.\citep{Gudiksen2002, Liao2007,Jie2008,Demichel2010,Calarco2011, Marquardt2013, Auzelle2020} Furthermore, the large aspect ratio of NWs also promotes the elastic relaxation of mismatch strain between the substrate and the NW as well as in NW heterostructures, resulting in coherent interfaces below a certain critical NW diameter.
\citep{Ertekin2005, Raychaudhuri2006, Goldthorpe2008} Finally, the large NW surface-to-volume ratio makes these nanostructures particularly sensitive to physical and chemical interactions with the environment. This inherent characteristic has been extensively exploited to fabricate NW-based gas,\citep{Wan2004, Ahn2008, Patsha2015} chemical \citep{Liu2004} and humidity sensors \citep{Kuang2007} as well as to develop cancer markers detectors.\citep{Zheng2005}

Another surface related phenomenon whose effects become evident at the nanoscale for large enough surface-to-volume ratios is surface stress.\citep{Martinez1990, Cammarata1994, Ibach1997} Surface stress is in some respect similar to the effect of Laplace pressure under the curved surface of a liquid, but is different in one essential aspect. While for a liquid surface energy and surface tension coincide (the surface energy can only be changed by the creation of a new surface), that is not the case for a solid. The difference between surface energy and surface tension for solid surfaces was considered by Gibbs and stems from the two possible ways of modifying the surface and its energy, namely, by creating a new surface or by stretching the existing one without changing the number of surface atoms.\citep{Shuttleworth1950,Herring1951,Cammarata1994,Diehm2012}
Surface stress has been deeply investigated in various metals in the form of nanoparticles,\citep{Wasserman1972, Diao2004, Mays1968, Wasserman1970, Wasserman1972, Wei2007} NWs,\citep{Berger1997, Jing2006} thin films \citep{Banerjee2003} and nanoporous layers,\citep{Kramer2004} as well as in oxide nanoparticles \citep{Zhou2001, Wu2004, Swamy2006, Bhowmik2006, Selbach2007, Wu2007, Hailstone2009, Ali2009, Chen2010, Diehm2012} and thin films.\citep{Kossoy2006}
Concerning semiconducting materials, surface stress has been investigated in quantum dots \citep{Meulenberg2004} and nanoparticles,\citep{Shreiber2006, Diehm2012} and also in Si NWs by means of computational techniques.\citep{Park2012} However, to the best of our knowledge, the effects of surface stress have never been experimentally observed in semiconducting NWs.

In this work, we present evidence of surface stress in nearly coalescence-free self-assembled GaN NWs grown by plasma-assisted molecular beam epitaxy (PA-MBE) on Ti films sputtered on Al$_{2}$O$_{3}(0001)$. GaN NWs directly grown on metallic substrates have recently emerged as a promising material system for the fabrication of electronic and optoelectronic devices.\citep{Calabrese2016, May2016, Zhao2016, Zhao2016b, Calabrese2017, Treeck2018, Ramesh2019, Ramesh2019b, May2019, Sun2019, Tyagi2019, Ramesh2020} Here, we exploit the unique combination of small NW lateral dimensions and negligible coalescence degrees in optimized GaN NW ensembles grown on Ti \citep{Treeck2018, Calabrese2019} to investigate the strain state of these semiconducting nanostructures. In contrast to previous studies reporting on coalescence-free GaN NW ensembles,\citep{Hersee2009, Brubaker2011, Choi2012, Hugues2013, Gacevic2015, Hetzl2017, DeSouza2017} the samples analyzed here contain GaN only in the form of NWs, i.e., neither does the substrate contain GaN nor does a parasitic GaN layer form in between the NWs, which simplifies strain analysis of the GaN NWs. The examination of different coalescence-free GaN NW ensembles by high-resolution X-ray diffraction (HR-XRD) reveals the presence of homogeneous strain that increases with decreasing of the average NW radius. This radius-dependent homogeneous strain is explained in terms of surface stress acting on the NW sidewalls, and becomes particularly relevant in ensembles of thin NWs. Additionally, the broadening of higher-order Bragg peaks in the HR-XRD profiles indicates the presence of significant  strain variations within the ensemble. This strain variations do not arise from fluctuations of the lattice constant within individual NWs, but from variations of the lattice constant from NW to NW due to the superposition of different homogeneous strain states in the NW ensemble, mimicking an inhomogeneous strain on a macroscopic scale. The superposition of diverse strain states also leads to a broadening of the excitonic transitions in continuous-wave photoluminescence (cw-PL) spectra, resulting in linewidths comparable to those reported for NW ensembles prepared on Si.\citep{Fernandez-Garrido2014} Finally, by combining X-ray diffraction and photoluminescence spectroscopy, the surface stress components $f_{x}$ and $f_{z}$  of the air-exposed GaN$\{1\bar100\}$ planes are experimentally determined.

\begin{table*}
\caption{\label{tab:1} Ga and N fluxes ($\Phi_{\mathrm{Ga}}$ and $\Phi_{\mathrm{N}}$), growth temperature $T$, growth time $t$, and average NW radius $R$ for all investigated samples.}
\begin{tabular}{ccccccc}
\hline
\hline
Sample & $\Phi_{\mathrm{Ga}}$ (ML/s) & $\Phi_{\mathrm{N}}$ (ML/s) & $T$\,($^{\circ}$C) & $t$ (min) & $R$ (nm) & \tabularnewline
\hline
A & 0.27 & 0.36 & 710 & 240 & 22 & \tabularnewline
B & 0.32 & 0.75 & 630 & 120 & 15 & \tabularnewline
C & 0.39 & 1.05 & 610 & 300 & 22 & \tabularnewline
D & 0.32 & 0.75 & 600 & 120 & 29 & \tabularnewline
E & 0.27 & 0.36 & 600 & 240 & 63 & \tabularnewline
\hline
\hline
\end{tabular}
\end{table*}

\section{Results and discussion}

\subsection{Morphology, coalescence degree, and orientation of GaN NWs grown
on Ti/Al$_\mathbf{2}$O$_\mathbf{3}\mathbf{(0001)}$}

\begin{figure*}
\includegraphics[width=0.9\textwidth]{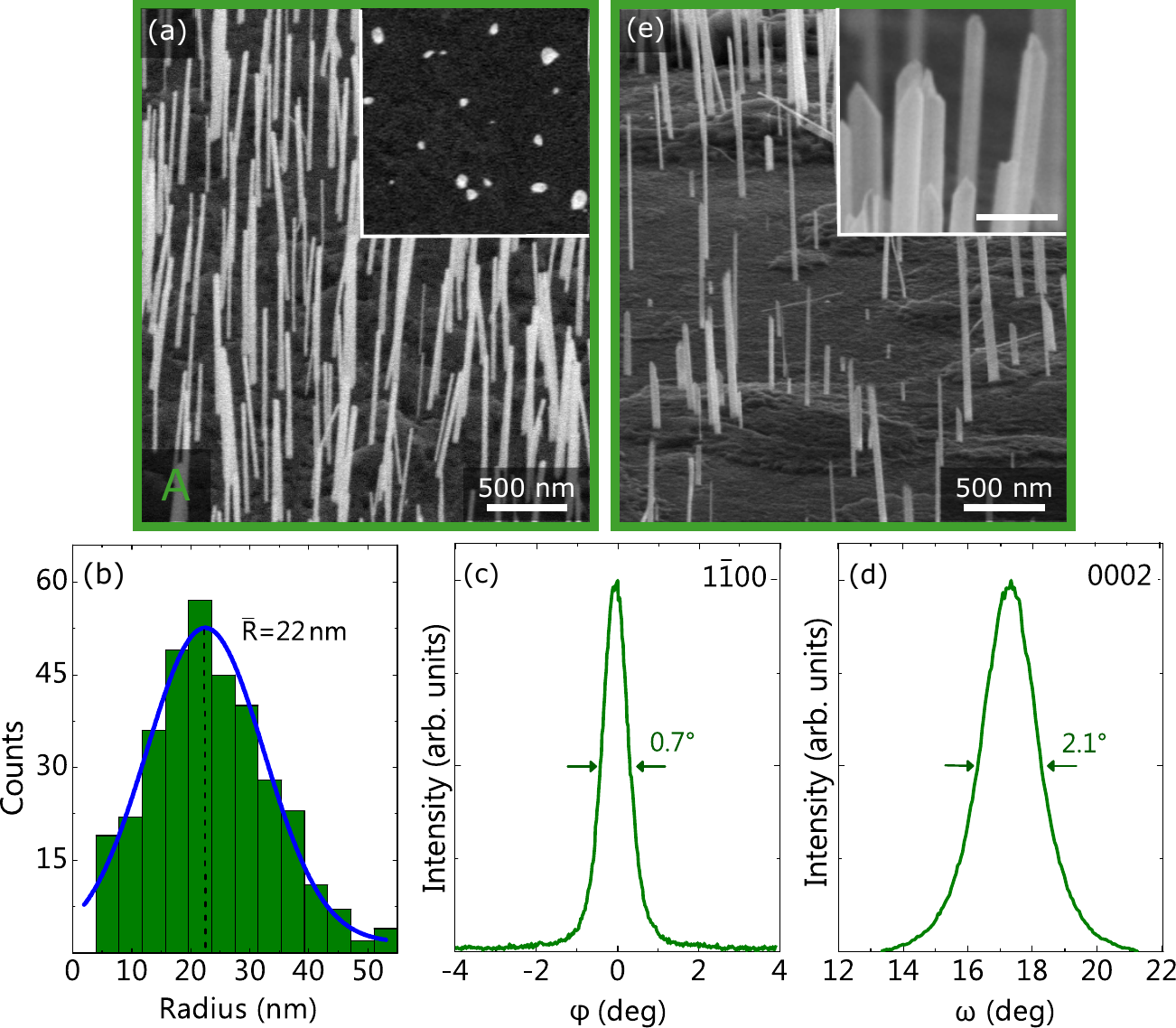} 
\caption{(a) Bird’s eye view scanning electron micrograph of sample A. The inset presents a top-view scanning electron micrograph of the same sample. (b) Distribution of the NW radii obtained from top-view scanning electron micrographs. The average NW radius is determined from the fit of a normal distribution to the experimental data, which is shown as a solid line. (c) and (d) XRD $\varphi$ and $\omega$ scans, respectively, across the GaN~$1\bar{1}00$ and $0002$ reflections of sample A. (e) Bird’s eye view scanning electron micrograph of sample A after 90 min etching in a 5M~KOH solution at 40\,$^{\circ}$C. The micrograph shown in the inset illustrates the pencil-like shape of etched NWs. The scale bar in the inset corresponds to 1~\textmu m. }
\label{SEMXRD} 
\end{figure*}

Due to the potential impact of the NW morphology, coalescence degree, and orientational distribution on the actual strain state of GaN NW ensembles, we first analyze all these aspects in a representative GaN NW sample grown on Ti/Al$_{2}$O$_{3}$. Figure~\ref{SEMXRD}(a) shows a bird’s-eye view scanning electron micrograph of sample~A. The growth conditions of this and all other samples studied in the present work are summarized in Table \ref{tab:1}. Despite an average NW length of about 2~\textmu m, the NW density of sample A is about 1$\times10^{9}$~cm$^{-2}$, almost one order of magnitude lower than the typical values reported for GaN NW ensembles grown on Si and on other types of substrates.\citep{Calarco2007,Schuster2012,Kumaresan2016} This result is attributed to the large Ga adatom diffusion length on TiN, which results in diffusion-induced repulsion of neighboring NWs.\citep{Treeck2018} As a consequence of the reduced NW number density, the NWs of sample A are virtually free of coalescence. The almost negligible degree of coalescence is clearly visible in the top-view scanning electron micrograph shown in the inset of Fig.~\ref{SEMXRD}(a). As can be observed, most NWs nucleate relatively far from each other. The actual coalescence degree of sample A, as derived from the manual inspection of several hundreds of NWs observed in top-view scanning electron micrographs, is 5\%.  This value is at least ten times lower than those of GaN NW ensembles grown on Si~with a comparable length.\citep{Brandt2014}

Figure \ref{SEMXRD}(b) presents the NW radius ($R$) distribution in sample A obtained from top-view scanning electron micrographs. A Gaussian fit returns a mean radius of $\bar{R}=22 $\,nm and a standard deviation of the radial distribution of $\Delta R = 10$~nm. 

The orientation distribution of the NWs is investigated by XRD. The in-plane orientation distribution within the NW ensemble (the twist) is assessed by a $\varphi$ scan across the GaN $1\bar{1}00$ reflection [Fig.~\ref{SEMXRD}(c)]. The full-width-at-half-maximum (FWHM) of this reflection evidences a twist of 0.7$^{\circ}$. This value is comparable to the ones previously reported for GaN NWs grown on AlN/6H-SiC$(000\overline{1})$ \citep{Fernandez-Garrido2014} as well as on epitaxial multilayer graphene,\citep{Fernandez-Garrido2017} and significantly smaller than those obtained in NW ensembles prepared on Si (2--3$^{\circ}$).\citep{Jenichen2011,Fernandez-Garrido2014}  The reduced twist, resulting from the strict epitaxial relationship existing between the wurtzite GaN NWs and the rocksalt TiN substrate [$\alpha$-GaN(0001)$\langle 11\bar{2}0\rangle\|\delta$-TiN(111)$\langle 1\bar{1}0\rangle$],\citep{Woelz2015,Treeck2018} is a consequence of the single-crystalline nature of $\delta$-TiN,\citep{Calabrese2019} which forms at the surface of the Ti sputtered film upon exposure to the N plasma.\citep{Woelz2015}

To assess the NW out-of-plane orientation distribution (the tilt), we record an $\omega$ scan across the GaN~$0002$ reflection {[}Fig.~\ref{SEMXRD}(d){]}. The FWHM of this reflection reveals a tilt of 2.1$^{\circ}$, a value that is similar to those of GaN NW ensembles grown on Si.\citep{Jenichen2011,Kaganer2012,Fernandez-Garrido2014}  The value of the tilt is, however, significantly larger than that on AlN/6H-SiC$(000\overline{1})$ \citep{Fernandez-Garrido2014} and epitaxial multilayer graphene \citep{Fernandez-Garrido2017} for which a tilt of 0.4$^{\circ}$ was observed.  We attribute the large tilt value to the combination of two distinct phenomena. First, the occurrence of strong interfacial reactions between the impinging Ga atoms and the metallic substrate lead to a roughening of the substrate surface that promotes the elongation of misoriented NWs.\citep{Calabrese2019} Secondly, the negligible degree of coalescence minimizes the reduction of the average tilt caused by the merging of misoriented NWs.\citep{Consonni2009}

Finally, we investigate the NW polarity on a macroscopic scale by KOH etching. For this purpose, a piece of the sample is etched in a 5M KOH aqueous solution at 40\,$^{\circ}$C for 90~min. The analysis of the etched sample by scanning electron microscopy [Fig.~\ref{SEMXRD}(e)] reveals that the NWs become shorter and their areal density decreases upon KOH exposure. Moreover, after the chemical etching, the NWs develop a pencil-like shape, as visible in the inset of Fig.~\ref{SEMXRD}(e). The etching of the NWs in KOH shows that they grow along the $[000\bar{1}]$ axis,\citep{Hestroffer2011,Romanyuk2015} i.\,e., they are N polar.

\subsection{Positions of $\mathbf{000L}$ Bragg peaks and homogeneous strain $\mathbf{\varepsilon_{zz}}$}

\begin{figure*}
\includegraphics[width=0.7\textwidth]{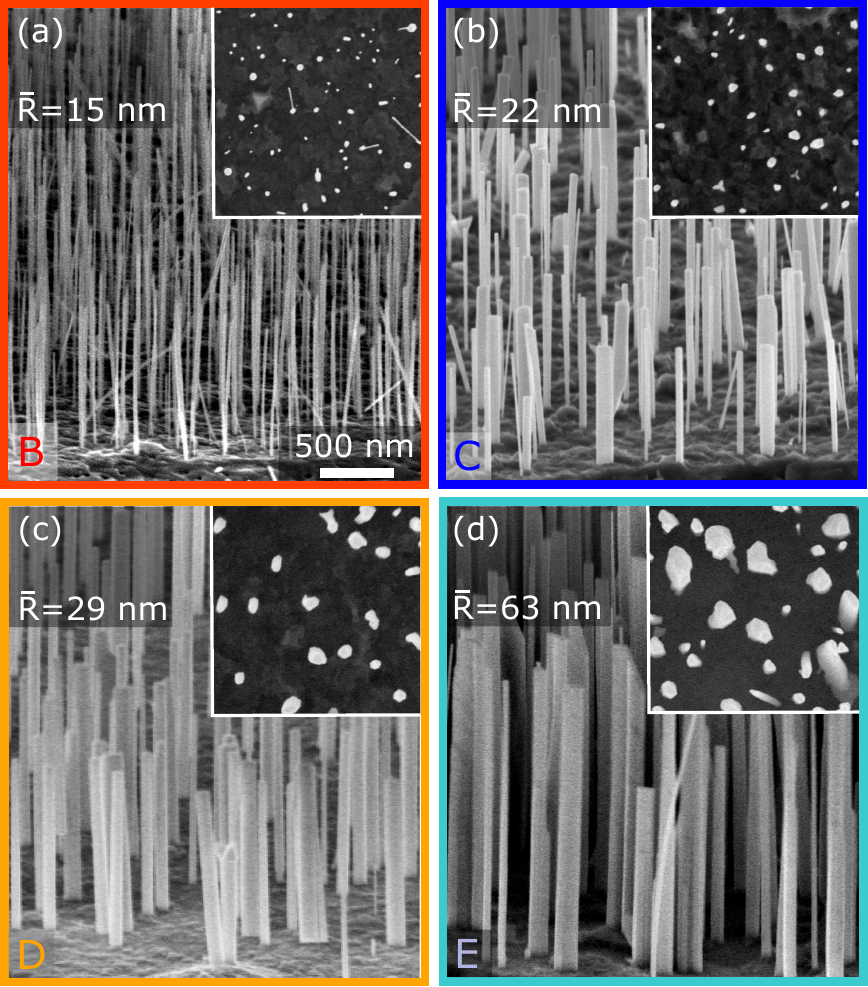} 
\caption{(a) Bird’s eye-view scanning electron micrographs of samples B--E. The insets are top-view scanning electron micrographs of the corresponding samples. The scale bar in (a) applies to all micrographs.}
\label{SEM}
\end{figure*}

In this section, we compare the angular positions of $000L$ Bragg peaks, sensitive to the presence of homogeneous strain in the NW axial direction, for NW ensembles of different radii. Figure \ref{SEM} presents exemplary bird's-eye and top-view scanning electron micrographs of samples B--E. In these samples the mean radius $\bar{R}$ varies between 15 and 63~nm (see Table~\ref{tab:1}) and the coalescence degree is small in all cases (it remains in the 5--9\% range for samples A--D and increases up to about 20\% for sample E).

\begin{figure*}
\includegraphics[width=0.95\textwidth]{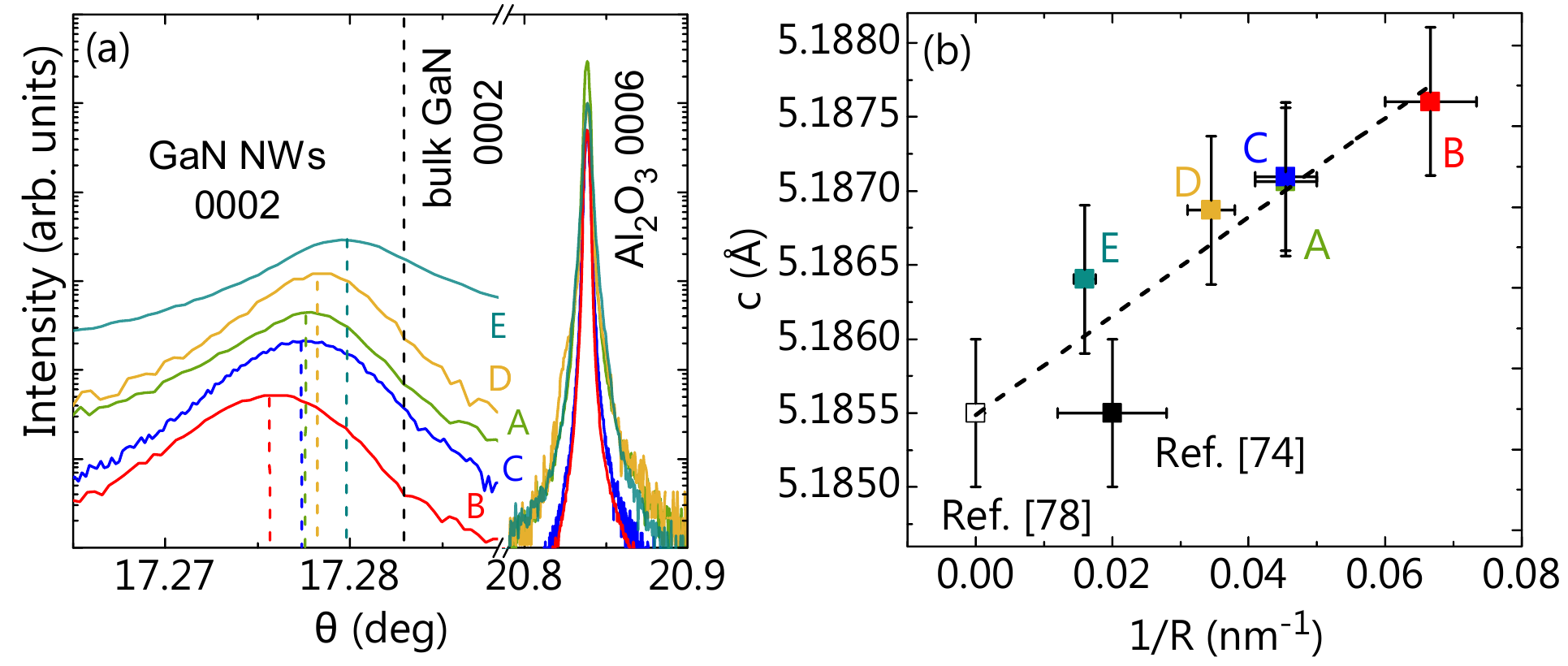}
\caption{(a) XRD $\theta/2\theta$ symmetric scans across the GaN~$0002$ and Al$_{2}$O$_{3}~0006$ reflections for samples A--E.  The profiles are normalized to the substrate peak and shifted vertically for clarity. For each sample, the position of the center of the GaN~$0002$ diffraction peak, which is determined from a Gaussian fit, is indicated by a dashed line. The expected peak position for bulk GaN derived from Ref.~\citenum{Moram2009} is also shown as a reference.  (b) Variation of the $c$ lattice constant as a function of $1/R$ for samples A--E. The $c$ lattice parameters of bulk GaN \citep{Moram2009} and of a GaN NW ensemble grown on Si (possessing a high coalescence degree) \citep{Jenichen2011} are also shown for comparison. The dashed straight line is the least-squares fit of the data.}
    \label{HOMOGENEOUS}
    \end{figure*}

Figure \ref{HOMOGENEOUS}(a) presents $\theta/2\theta$ scans across the GaN~$0002$ and the Al$_{2}$O$_{3}~0006$ reflections for samples A--E. The peak positions of the GaN~$0002$ reflection differs from sample to sample, indicating that the NW ensembles possess different average \textit{c} lattice parameters.  The position of the Bragg angle for the $0002$ reflection of a bulk GaN crystal derived from the average value of the $c$ lattice constant of bulk GaN reported in Ref.~\citenum{Moram2009} ($c_{\mathrm{GaN}}=5.1855$~$\textrm{Å}$, see also Ref.~\citenum{Darakchieva2007} for further details) is indicated in Fig.~\ref{HOMOGENEOUS}(a) by a vertical dashed line at $\theta$= $17.2829^{\circ}$.  For all the investigated samples, the position of the GaN~$0002$ reflection is found at Bragg angles smaller than the one of bulk GaN, revealing that uncoalesced GaN NWs on Ti/Al$_{2}$O$_{3}$ are subjected to a net out-of-plane dilatation.

The absolute values of the \textit{c} lattice constants for samples A--E are determined using the Al$_{2}$O$_{3}~0006$ reflection as a reference, assuming that the substrate is free of strain and taking the \textit{c}-Al$_{2}$O$_{3}$ lattice constant equal to 12.9920~Å. The slightly broader substrate peaks observed for samples D and E are attributed to residual inhomogeneous strain in the substrate introduced by the sputtered Ti film. The values of the \textit{c} lattice parameter derived from the position of the GaN~$0002$ reflection from samples A--E are presented in Fig.\,\ref{HOMOGENEOUS}(b) as a function of the inverse radius. 
The $c$ lattice parameters of bulk GaN and of a GaN NW ensemble grown on Si, which are retrieved from Refs.~\citenum{Moram2009} and \citenum{Jenichen2011}, respectively, are included for comparison. The error bars correspond to an uncertainty of $\pm 5\times10^{-4}$~Å in the values of $c$ and of 10\% for $R$, a value determined from the analysis of top-view scanning electron micrographs. The $c$ lattice constants for the different NW ensembles investigated here are found to linearly increase with $1/R$. Hence, the measurements reveal a homogeneous strain $\varepsilon_{zz}=\Delta c/c$ in ensembles of uncoalesced GaN NWs on Ti/Al$_{2}$O$_{3}$ exhibiting sub-50~nm radii that depends on $R$ as

\begin{equation}
\varepsilon_{zz}=\alpha/R,\label{eq:epszz}
\end{equation}
where $\alpha=(6.4 \pm 1.3)\times10^{-3}$\,nm.

This radius-dependent homogeneous strain in GaN NWs has not been reported up to now. Homogeneous strain in GaN NWs was only reported in ensembles of closely spaced GaN NWs \citep{Auzelle2016} and attributed to the high degree of coalescence associated to dense NW ensembles. In the present case, the observed radius-dependent homogeneous strain is clearly not the result of NW coalescence since the samples under investigation are nearly coalescence-free.  While the combination of the large lattice constant and thermal expansion coefficient mismatches (6.2\% and 40.2\%, respectively) existing between GaN and $\delta$-TiN might introduce in-plane compressive strain and out-of-plane tensile strain in the GaN NWs at the interface, epitaxial strain in NWs exponentially decays along the NW length within a characteristic distance equal to $R$.\citep{Kaganer2012,Hugues2013}  Also, if the lattice mismatch would be the origin of the observed homogeneous strain, thicker NWs (sample D) would give rise to larger strain contrary to our experimental observations (see Fig.~\ref{HOMOGENEOUS}). The source of homogeneous strain in our ensembles of thin and uncoalesced GaN NWs must thus be a different one, as further discussed below.

\subsection{Broadening of Bragg peaks and strain inhomogeneity}

\begin{figure}
\includegraphics[width=0.9\columnwidth]{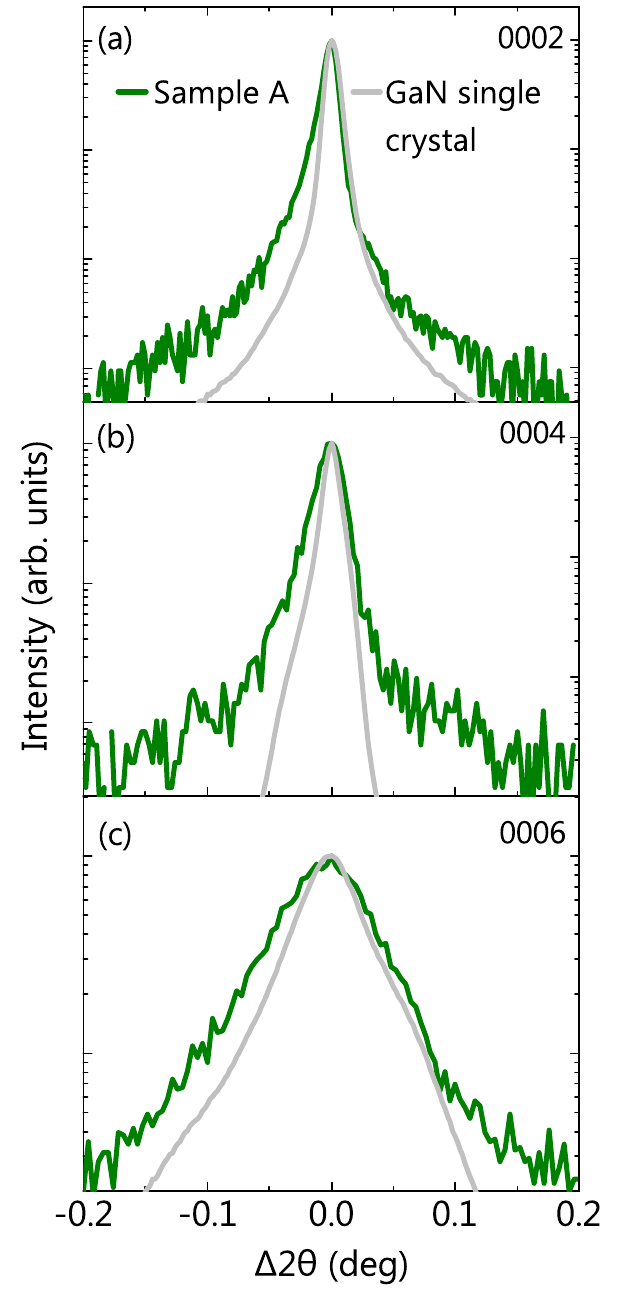} 
\caption{(a) XRD $\theta/2\theta$ scans along the symmetric GaN (a) $0002$, (b) $0004$ and (c) $0006$ Bragg reflections for sample A (green curves) and the bulk GaN crystal taken as reference (grey curves).} \label{INHOMOGENEOUS}
\end{figure}

Inhomogeneous strain gives rise to lattice spacing variations that can be detected by XRD as a broadening of the $\theta/2\theta$ scans. $\theta/2\theta$ scans across the GaN~$0002$, $0004$ and $0006$ reflections from sample~A are shown in Fig.~\ref{INHOMOGENEOUS} together with identical measurements performed on the bulk GaN reference sample. Since our analysis is restricted to symmetric Bragg reflections, only the strain component along the NW axis is involved in the study. The diffraction profiles are asymmetric for both sample~A and the bulk reference sample. This asymmetry is a direct consequence of the vertical (perpendicular to the scattering plane) divergence of the X-ray beam, as explained in Ref.~\citenum{Kaganer2012}. For the $0002$ reflection, the FWHM of the diffraction peak for sample~A is almost the same as for the reference sample {[}Fig.~\ref{INHOMOGENEOUS}(a){]}.  An enhanced diffuse scattering from sample A at large deviation angles from the position of the Bragg peak is attributed to the broad out-of-plane orientational distribution of the NW ensemble.  On the other hand, the GaN $0004$ and $0006$ reflections from sample~A are broader than those from bulk GaN {[}Figs.~\ref{INHOMOGENEOUS} (b) and~\ref{INHOMOGENEOUS}~(c){]}.

\begin{figure}
\includegraphics[width=\columnwidth]{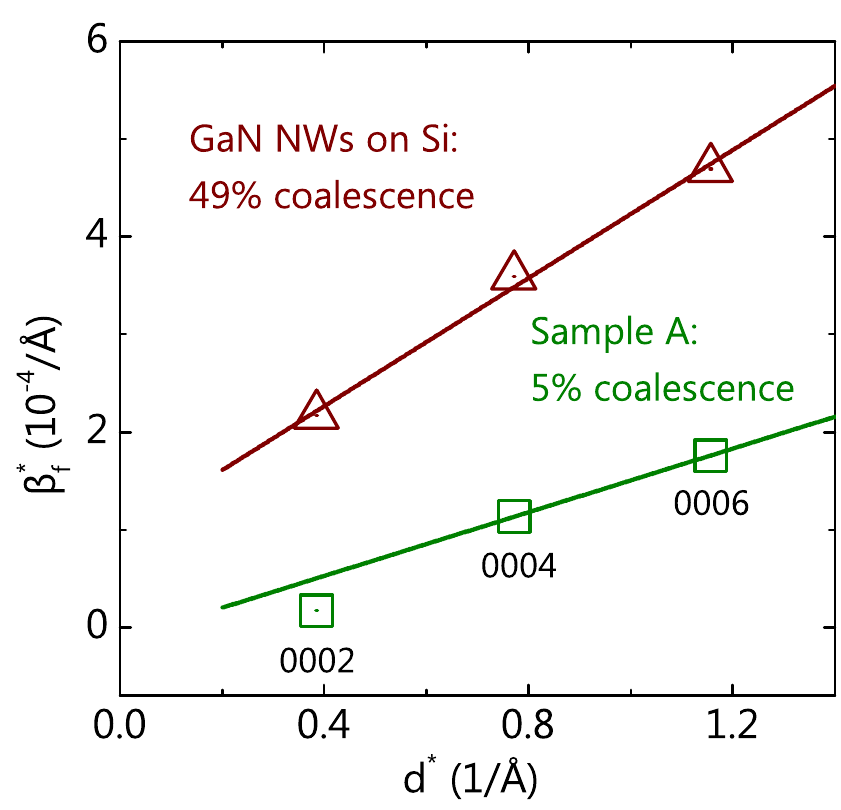} 
\caption{Williamson-Hall plot for the symmetric GaN reflections of sample A and of a GaN NW ensemble grown on Si.\citep{Fernandez-Garrido2014} The straight lines are linear fits to the experimental data.}
\label{WH} 
\end{figure}

Since the strain broadening of the X-ray diffraction peaks is proportional to the reflection order, the root mean square (rms) strain variation is commonly derived from a Williamson-Hall (WH) plot.\citep{Williamson1953} Figure \ref{WH} shows the WH plot in the reciprocal space representation given by $\beta_{f}^{*}=(\beta_{f}\cos\theta)/\lambda$ versus the inverse lattice spacing $d^{*}=(2\sin\theta)/\lambda$, where $\theta$ is the Bragg angle, and $\beta_{f}$ is the integral breadth of the diffraction profile corrected by the breadth of the resolution function:
\begin{equation}
\label{WHeq}
\beta_{f}^{*}=\beta_{s}^{*}+2 e d^{*}.
\end{equation}
Here, $\beta_{s}^{*}$ is the broadening of the diffraction peak due to the finite NW length, and the rms strain variation is $(2/\pi) e$ because a Lorentzian is used here to fit the diffraction profiles.

Figure \ref{WH} compares the WH plot for sample A with the measurements on a GaN NW ensemble grown on Si$(111)$ whose fabrication details are reported elsewhere.\citep{Fernandez-Garrido2014} The latter has a coalescence degree of 49$\%$, notably smaller than typical values reported for GaN NWs on Si of about 80$\%$.\citep{Brandt2014, Fernandez-Garrido2014} From the linear fits, we obtain slopes of 1.6$\times10^{-4}$ and 3.3$\times10^{-4}$ for sample A and the reference GaN NW ensemble, respectively. The slope for sample A is obtained from the 0004 and 0006 reflections since the integral breadth of the 0002 reflection of sample A, corrected for the instrumental resolution, is so close to zero that it would result in a negative intercept. From the slopes obtained by the fits, we find an rms strain of 5.2$\times10^{-5}$ for sample A and a twice larger value of 1.0$\times10^{-4}$ for the reference sample on Si.

At a first glance, the strain broadening of the Bragg peaks of the NW ensemble on Ti is surprisingly large given that the coalescence degree for sample A is one order of magnitude smaller as compared to that of the reference sample on Si. 
However, the radius-dependent homogeneous strain revealed above, together with the broad distribution of the NW radii, will inevitably result in a broadening of the reflections in ensemble measurements. We can easily estimate the Bragg peak width resulting from this superposition of different homogeneous strain states in the ensemble. Using Eq.\,(\ref{eq:epszz}), we estimate the lattice spacing variation over the ensemble of NWs as $\Delta d/d=\Delta(\alpha/R)=(\alpha/\bar{R})(\Delta R/\bar{R})$, where $\Delta R$ is the mean-squared radius deviation from the average radius $\bar R$. With the values $\bar{R}=22$\,nm and $\Delta R=10$\,nm for sample A, we find an apparent rms strain of $1.2\times10^{-4}$, a value even larger than the one determined from the analysis of the WH plot. This result implies that the broadening of the Bragg peaks for sample A is in fact entirely due to the radius-dependent homogeneous strain, and has no contribution from actual strain fluctuations within single NWs, such as induced when NWs coalesce by bundling.\citep{Kaganer2012,Kaganer2016}

\subsection{Energy and linewidth of excitonic transitions in PL spectra}

\begin{figure}
\includegraphics[width=0.9\columnwidth]{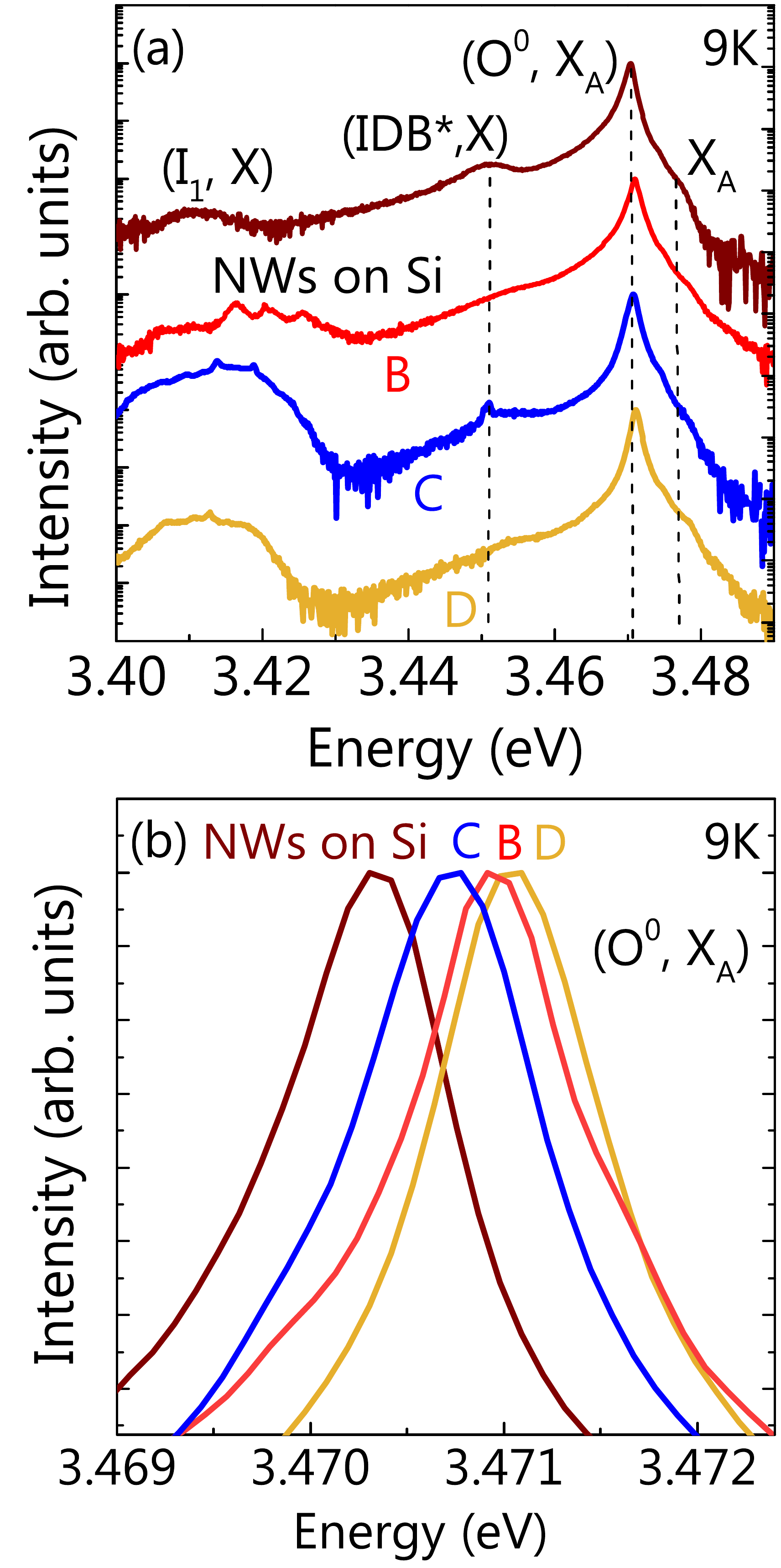} 
\caption{(a) Low temperature (9 K) PL spectra of samples B--D on logarithmic scale.  The spectrum of the reference GaN NW ensemble prepared on Si$(111)$ is also included for comparison. The spectra are normalized and vertically shifted for clarity. (b) Same spectra as shown in (a) but on linear scale and restricted to the (O$^{0},X_{A})$ transition. } \label{PL} 
\end{figure}

The homogeneous strain observed in samples A--D is expected to shift not only the position of the diffraction peaks in XRD $\theta/2\theta$ scans, but also the position of donor-bound exciton transitions in PL experiments. To check this expectation, near-band-edge PL spectra for samples A--D are collected at 9~K. Figure~\ref{PL}(a) shows the normalized PL spectra of samples B--D together with the one of a reference GaN NW ensemble grown on Si at 835\,$^{\circ}$C,\citep{Zettler2015} as measured by an optical pyrometer.

The PL spectra of samples A and E are not reported here because they are affected by diffusion of O from the Al$_{2}$O$_{3}$(0001) substrate and its incorporation into GaN.\citep{Calabrese2019} This effect is observed only for these two samples because of the reduced Ti film thickness, which promotes O interdiffusion into GaN.\citep{Calabrese2019} However, the width and energy of the near-band edge transition in the PL spectrum of samples A and E indicate an O concentration not higher than $10^{19}$ cm$^{-3}$,\citep{Feneberg2014} which translates into an O induced hydrostatic strain of about 2$\times10^{-5}$.\citep{Walle2003}  Since this value is one order of magnitude smaller than the strain measured by XRD [see Fig.\,\ref{HOMOGENEOUS}(b)], samples A and E are still suitable for the analysis of the homogeneous strain as a function of the average value of $R$.

The PL spectra of samples B--D are dominated by the recombination of excitons bound to neutral oxygen donors (O$^{0},X_{A})$ at about $3.47$~eV, in agreement with previous work.\citep{Corfdir2014} We also observe in all samples, between $3.40$ and $3.43$~eV, the radiative recombination of excitons bound to $I_{1}$ basal-plane stacking faults $(I_{1},X_{A})$.\citep{Corfdir2014b} Besides these transitions, for sample C and the NW ensemble grown on Si, we additionally detect the recombination of excitons bound to inversion domain boundaries (IDB$^{*},X)$ at $3.45$~eV.\citep{Auzelle2015, Pfuller2016} A closer look at the (O$^{0},X_{A})$ transition shows that for samples B--D this line is always slightly blue-shifted as compared to the spectrum of the reference sample, for which the transition is centered at $3.4703$~eV {[}see Fig.~\ref{PL}(b){]}.

Surprisingly, despite the absence of coalescence, the full width at half maximum of the (O$^{0},X_{A})$ line (FWHM$_{\mathrm{PL}}$) for the NW ensembles grown on Ti/Al$_{2}$O$_{3}$ are comparable to the value of the reference sample prepared on Si {[}see Fig.~\ref{PL}(b){]}. This result is attributed to two distinct effects. First, for an average NW radius of 20~nm, about one half of the donors are located at a distance of less than one Bohr radius from the surface. The resulting energy dispersion of donor-bound excitons cause a broadening on the order of 1~meV.\citep{Brandt2010,Corfdir2014} Second, a broadening of the PL lines is also induced by the distribution of radii in the NW ensembles and the resulting distribution of transition energies, as a consequence of the strain dependence on the radius. The magnitude of this effect can be estimated from the positions of the PL lines in Fig.~\ref{PL} (a). Since the widths of the radial distributions are comparable with the mean radii, the widths of the PL lines are expected to be comparable with the shifts of the PL lines with respect to the reference sample, i.e., again on the order of 1~meV. As a consequence of the combination of these two effects, the PL linewidth is comparable to the values observed for ensembles of thicker NWs on Si (1--2~meV), for which a part of the broadening is due to coalescence-induced strain variation. We point out that the PL energy position of samples B--D is unlikely to be affected by the dielectric confinement of excitons, which was reported to cause a blueshift of the PL lines only for NWs with even smaller diameters as the present ones.\citep{Zettler2016}

\subsection{Effect of surface stress}

\begin{figure}
\includegraphics[width=1\columnwidth]{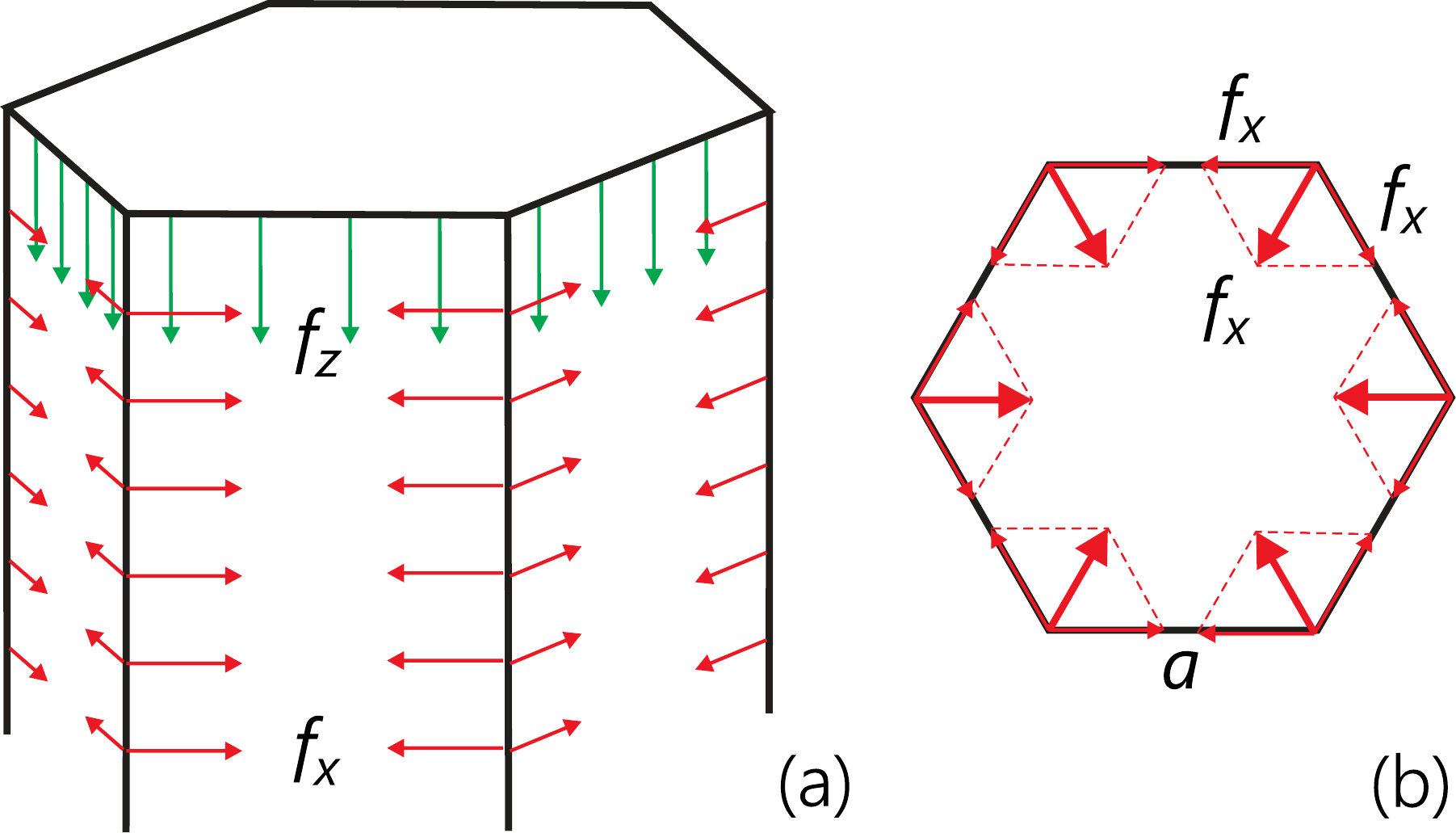} 
\caption{(a) Sketch illustrating the surface stress acting on a faceted GaN NW: the force densities $f_{z}$ and $f_{x}$ are applied at the NW top and side edges, respectively. Both force densities result in a net bulk strain within the NW. (b) Further illustration of the force density $f_{x}$ acting on a GaN NW, as seen in cross-sectional view.}
\label{SurfaceStress}
\end{figure}

The \textit{c} lattice parameter variation of the GaN NWs in samples A--E is found to be inversely proportional to the NW radius, as shown in Fig.~\ref{HOMOGENEOUS}(b).
We can exclude explanations of this phenomenon specific to ionic systems.\citep{Diehm2012} In the following, we show that the most plausible origin of the observed radius-dependent homogeneous strain in GaN NWs is surface stress.
Surface stress is a tensor quantity, reflecting the fact that the surface, depending on its symmetry, responds differently to stretching along different directions. The sidewalls of GaN NWs are $\{1\bar100\}$ (\textit{M}-plane) facets, and for symmetry reasons the shear components of the surface stress tensor are absent. Hence, the surface stress tensor possesses only two components, $f_{z}$ and $f_{x}$. They represent forces per unit length acting along the $\langle 0001 \rangle$ and $\langle 11\bar{2}0 \rangle $ directions between the atoms at the surface of a facet. These forces are compensated within the facets but not at their intersections, giving rise to net forces applied at the edges of the crystal.\citep{Herring1951} This effect is sketched in Figs.~\ref{SurfaceStress}(a) and \ref{SurfaceStress}(b), which show a schematic diagram of an individual NW having hexagonal cross-sectional shape and the linear force densities $f_{x}$ and $f_{z}$ acting on it.

Since the in-plane forces $f_{x}$ at the edges between facets form an angle of 120$^{\circ}$ with respect to each other, their sum is a vector of length $f_{x}$ pointing radially. These linear force densities introduce an in-plane strain in each NW, in a similar way to the effect of the Laplace pressure under the curved surface of a liquid. Strictly speaking, the strain produced by these forces is inhomogeneous because they are applied at the edges of the hexagonal NW rather than being homogeneously distributed over the whole NW external surfaces. However, the strain state differs rather little from a homogeneous strain, because the hexagonal NW cross-section differs only slightly from a circle.

The linear force densities $f_{z}$ are applied to the perimeter of the top facet, rather than uniformly over its area [see Fig.~\ref{SurfaceStress}(a)].  In accordance with Saint-Venant's principle, the stress within a single NW remains inhomogeneous only at distances from the top facet smaller than the NW diameter, so that the whole NW, except its uppermost part, is expected to be homogeneously strained as a consequence of $f_{z}$.

In the following, we neglect the strain inhomogeneity due to the NW edges. In addition, for simplicity, we consider that the NWs instead of having a hexagonal cross-section with side $a$, have a cross-section with the same area but with a circular shape. Therefore, if $R$ is the NW radius, the cross-sectional area of our cylindrical NWs is given by $\pi R^{2}=(3\sqrt{3}/2)a^{2}$, so that $R\approx0.91a$. The six linear force densities $f_{x}$ acting at the side edges of a hexagonal NW can thus be replaced by the areal force density $f_{x}/a$, which is homogeneously applied to the side surface of the cylinder.  This areal force density results in the bulk stress components $\sigma_{rr}=\sigma_{\theta \theta}=-f_{x}/a$. Here, we proceed to cylindrical coordinates, assuming axial symmetry of the problem, so that $\sigma_{xx}=\sigma_{yy}=\sigma_{rr}=\sigma_{\theta \theta}$ and $\varepsilon_{xx}=\varepsilon_{yy}=\varepsilon_{rr}=\varepsilon_{\theta \theta}$.

At the top facet, the force with the linear density $f_{z}$ is applied along the cross-sectional perimeter {[}see Fig.~\ref{SurfaceStress}(a){]}. Under the cylindrical approximation, the NW cross-sectional perimeter has a length of 6$a$. Consequently, the linear density force $f_{z}$ divided by the area of the top facet, $(3\sqrt{3}/2)a^{2}$, results in the bulk stress component $\sigma_{zz}=-(4/\sqrt{3})f_{z}/a$.  All components of the bulk stress are thus inversely proportional to $a$, and can be related to the strain components, $\varepsilon_{ij}$, using Hooke's law,
\begin{align}
\sigma_{rr} & = c_{11}\varepsilon_{rr}+c_{12}\varepsilon_{\theta\theta}+c_{13}\varepsilon_{zz},\nonumber \\
\sigma_{\theta\theta} & = c_{12}\varepsilon_{rr}+c_{11}\varepsilon_{\theta\theta}+c_{13}\varepsilon_{zz},\label{eq:Hookes} \\
\sigma_{zz} & = c_{13}(\varepsilon_{rr}+\varepsilon_{\theta\theta})+c_{33}\varepsilon_{zz},
\nonumber
\end{align}
where $c_{ij}$ are the elastic moduli of GaN.

This strain will induce a variation in the bandgap energy, which in the first approximation is proportional to strain:
\begin{equation}
\Delta E=a_{\mathrm{PL}}(\varepsilon_{rr}+\varepsilon_{\theta\theta})+b_{\mathrm{PL}}\varepsilon_{zz},
\label{deformation_potentials}
\end{equation}
where $a_{\mathrm{PL}}$ and $b_{\mathrm{PL}}$ are the deformation potentials, with values of $-7.1$ and $-11.3$~eV, respectively.\citep{Kaganer2015}

According to the data in Fig.~\ref{HOMOGENEOUS}(b), the strain $\varepsilon_{zz}=\Delta c/c$ with respect to the $c$ lattice parameter of bulk GaN varies from $2.4\times 10^{-4}$ for sample D to $4\times 10^{-4}$ for sample B. Then, the term $b_\mathrm{PL}\varepsilon_{zz}$ would result in a redshift of the PL lines ranging from $-2.8$ to $-4.4$~meV, respectively. What we actually observe is a blueshift by about $+0.7$~meV [see Fig.~\ref{PL}(b)] with respect to the reference NW ensemble grown on Si, which is free from homogeneous strain.\citep{Jenichen2011} Hence, the effect of the out-of-plane strain is overcompensated by in-plane strain. A full compensation would be achieved at $\varepsilon_{zz}/\varepsilon_{rr}=-2a_\mathrm{PL}/b_\mathrm{PL}=-1.25$. Note that this strain state is very different from that induced by the Poisson effect in a biaxially strained GaN layer, for which $\varepsilon_{zz}/\varepsilon_{rr}=-2c_{13}/2c_{33}=-0.53$.
Since the shifts of the PL lines in Fig.\,\ref{PL} are notably smaller than the term $b_\mathrm{PL}\varepsilon_{zz}$, and do not show a $1/R$ dependence on the mean radius, we neglect the shift of the PL lines and take $\Delta E=0$ in Eq.\,(\ref{deformation_potentials}). Then, combining Eqs.\, (\ref{eq:epszz}), (\ref{eq:Hookes}), and (\ref{deformation_potentials}), we arrive at the surface stress components $f_{x}=2.25$~N/m and $f_{z}=-0.7$~N/m. Density functional theory (DFT) calculations \citep{Diehm2012} give $f_{x}=-0.106$~eV/\AA$^2=-1.7$~N/m and $f_{z}=-0.055$~eV/\AA$^2=-0.5$~N/m for these GaN facets, i.\,e., $f_x$ is of the opposite sign as the one determined here experimentally. We note that the DFT calculations are performed for an atomically clean facet in vacuum, while our study is performed on NWs exposed to air, which is expected to result in the formation of a thin GaO$_{x}$ oxide layer on the NW surfaces. In addition, this oxide layer will be covered by surface adsorbates such as water.

\subsection{Grazing incidence diffraction measurements of the in-plane strain}

\begin{figure*}
\includegraphics[width=\textwidth]{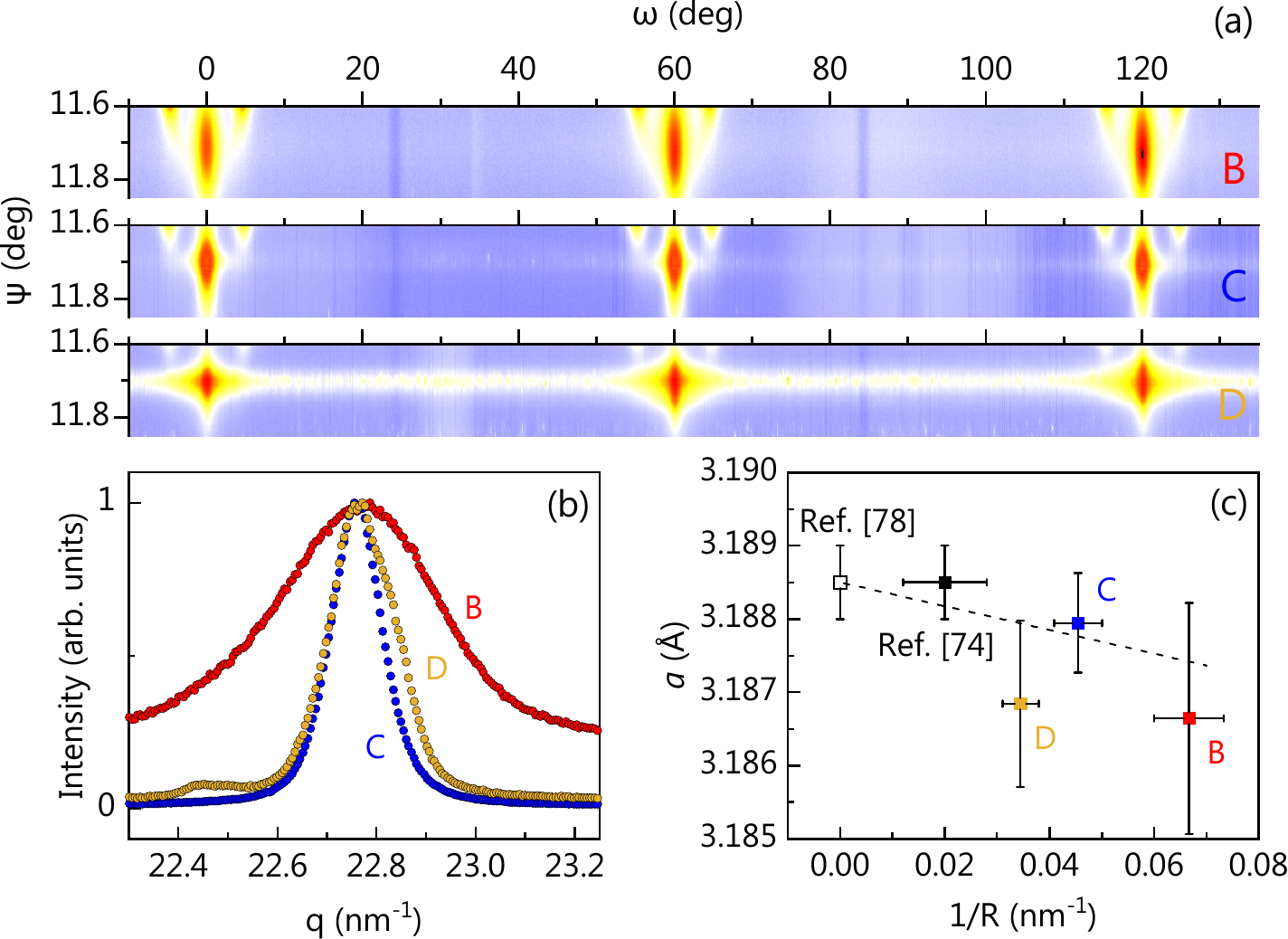} 
\caption{(a) GID $\omega-\psi$ maps of GaN~$1\bar100$ reflections of samples B--D obtained at a grazing incidence angle $\alpha=0.3^\circ$ and a grazing exit angle $\beta=0.15^\circ$, (b) profiles of the GaN~$1\bar100$ reflections at  $\alpha=0.45^\circ$ and $\beta=0.15^\circ$, and (c) the $a$ lattice parameters of GaN NWs of samples B--D obtained from the Bragg peak positions. The $a$ lattice parameters of bulk GaN \citep{Moram2009} and GaN nanowires on Si \citep{Jenichen2011} are shown for the comparison. The dashed straight line is the expected variation of the $a$ lattice parameter calculated on the basis of the XRD GaN~$0002$ data in Fig.\,\ref{HOMOGENEOUS} and PL data in Fig.\,\ref{PL}.}
\label{GID}  
\end{figure*}

For a comprehensive description of the strain state of samples B--D, we measure the in-plane lattice constants of GaN NWs and compare them with the ones previously determined by the combination of out-of-plane XRD and PL measurements. For this purpose, X-ray reciprocal space maps are collected around the GaN~$1\bar100$ reflections of samples B--D [see Fig.\,\ref{GID}(a)]. Each map is obtained by rotating the sample under investigation by an angle $\omega$ about the substrate surface normal. Reflections are repeated every $60^\circ$, in accordance with the hexagonal symmetry of GaN. The scans of intensity as a function of the scattering angle $\psi$, at the sample orientation $\omega$ corresponding to the maximum intensity, are used for the determination of the in-plane lattice spacing $a$. 

For the analysis of the diffracted intensity, we consider that the scattering plane, defined by the directions of the incident and the diffracted waves, is tilted with respect to the substrate surface, because of the finite grazing incidence and exit angles of the X-ray beam. The diffracted intensity from each NW, which is a thin long rod, is concentrated in the plane normal to the rod axis, in the direction of the GaN~$1\bar100$ reflections. The main contribution to the diffracted intensity comes from NWs strictly perpendicular to the scattering plane, while regions with smaller intensity are attributed to tilted and twisted NWs. 

The diffraction vector $\mathbf{q}$ has a component parallel to the plane of the substrate surface $q_{||}$, and a component, $q_z$, normal to it. These components can be expressed through the grazing incidence angle $\alpha$, the grazing exit angle $\beta$, and the scattering angle in the surface plane $\psi$ as
\begin{align}
q_{||}&=k_0\sqrt{\cos^2\alpha+\cos^2\beta-2\cos\alpha\cos\beta\cos\psi}, \nonumber \\
q_z&=k_0(\sin\alpha+\sin\beta), 
\label{eq:q}
\end{align} 

where $k_0=2\pi/\lambda$ is the wavevector. The total length of the scattering vector $q=(q_{||}^2+q_z^2)^{1/2}$ at the maximum of the diffracted intensity of a GaN~$1\bar100$ reflection determines the $a$ lattice spacing as $a=4\pi/q\sqrt3$. 

Figure \ref{GID}(b) presents the  GaN~$1\bar100$ X-ray intensity profiles of samples B--D obtained at an incidence angle $\alpha=0.45^\circ$ and an exit angle $\beta=0.15^\circ$. They are obtained, as a function of the scattering angle $\psi$, from reciprocal space maps analogous to those reported in Fig.\,\ref{GID}(a) and presented as a function of $q$, as explained above. Despite the broadening of the profiles due to the small cross-sectional size of NWs and their orientational distribution, one can see that the center of the diffraction peak of sample B is at larger wavevector as compared to the ones of samples C and D, reflecting a smaller in-plane lattice spacing.

Figure \ref{GID}(c) compares the $a$ spacings of samples B--D, obtained from the grazing incidence diffraction (GID) measurements described above, with the lattice parameters of bulk GaN \citep{Moram2009} and GaN NWs on Si.\citep{Jenichen2011} The $a$ lattice spacings are plotted versus $1/R$, as previously done in Fig.\,\ref{HOMOGENEOUS}(b) for the $c$ lattice spacing. We note that the lattice parameter of GaN nanowires on Si \citep{Jenichen2011} has been measured in a laboratory X-ray GID experiment basically similar to the present one. This measurement was possible because of a large NW density, providing enough diffracting volume for a laboratory experiment.

The error bars for samples B--D in Fig.\,\ref{GID}(c) are a result of the average over the measurements performed at different incidence and exit angles. We find out that the accuracy of the GID measurements, performed without an analyzer crystal, are insufficient for a reliable independent determination of the $a$ lattice spacing. However, the results are consistent with the $a$ spacing dependence expected from the $c$ lattice parameter measurements and the PL data, as shown by the dashed line in Fig.\,\ref{GID}(c). Hence, the contributions to the PL line shift from the lattice expansion along the GaN~$[0001]$ direction and the lattice contraction in the plane perpendicular to this direction compensate each other. 

\section{Summary and conclusions}

We have synthesized ensembles of GaN NWs on Ti characterized by different average NW radii and a negligible degree of coalescence, and observe that the out-of-plane strain component is inversely proportional to the mean radius of the NWs. We have shown that this phenomenon is due to surface stress acting on the NW sidewalls, which is only noticeable in ensembles of NWs possessing sufficiently small diameters. Despite the small diameters and the absence of coalescence, NW ensembles on Ti show a broadening of Bragg peaks which is only two times smaller than that detected in analogous structures with a high degree of coalescence grown on Si.  This effect is not caused by inhomogeneous strain within single NWs, as observed for NWs on Si, but due to the superposition of different homogeneous strain states in the NW ensemble, as a result of the radius-dependent strain and the broad NW radius distribution. 

The measured shift of the PL lines has been found much smaller than expected from the action of the measured strain $\varepsilon_{zz}$, and possesses an opposite sign. This result is an effect of GaN lattice contraction in the NW basal plane, in agreement with GID data. The accuracy of the GID measurements, performed without an analyzer crystal, was not sufficient for a direct determination of the surface stress components, but compatible with the values obtained by combining the $\varepsilon_{zz}$ strain component derived from HR-XRD with the PL line shift. Following this approach, we estimate that the surface stress components $f_{x}$ and $f_{z}$ are $2.25$ and $-0.7$~N/m, respectively. These values differ from those previously obtained by DFT calculations for clean \textit{M}-plane GaN. We attribute this difference to the oxidation of the NW surfaces and/or to the presence of adsorbates due to ambient exposure.

Despite the small strain inhomogeneity in GaN NWs grown on Ti/Al$_{2}$O$_{3}$, the linewidth of excitonic transitions associated to neutral donors in PL experiments is found to be comparable to those of ensembles of coalesced NWs. This result is caused by both the broad energy distribution of donors due to their varying distances to the NW sidewalls, and to a broad NW radius distribution.  Each effect contributes with about 1~meV to the broadening of the excitonic transitions.

\section{Methods}

\subsection{Growth}

Five different GaN NW ensembles (samples A--E) grown by PA-MBE on Ti films sputtered on Al$_{2}$O$_{3}$(0001) are investigated in this study. Before NW growth, a Ti film with a thickness of either 1.3~\textmu m (samples A and E) or 3.4~\textmu m (samples B--D) is deposited on bare Al$_{2}$O$_{3}$(0001) by magnetron sputtering as described elsewhere.\citep{Treeck2018}  After Ti sputtering, the samples are loaded into the growth chamber of our PA-MBE system, being exposed to air during this process. Our MBE system is equipped with a solid-source effusion cell for Ga and a radio-frequency N$_{2}$ plasma source for active N. The impinging Ga and N fluxes ($\Phi_{\mathrm{Ga}}$ and $\Phi_{\mathrm{N}}$, respectively) are calibrated in equivalent growth rate units of two-dimensional GaN(0001) layers \citep{Heying2000} and are expressed in monolayers per second (ML/s). The substrate temperature during NW growth is measured with a thermocouple placed in contact with the substrate heater.

Samples A--E differ in the thickness of the sputtered Ti film, as indicated above, the substrate nitridation step preceding the formation of the GaN NWs, and the GaN growth conditions. The GaN growth conditions involve $\Phi_{\mathrm{Ga}}$ and $\Phi_{\mathrm{N}}$, the substrate temperature, and the growth time. For samples A, D and E, the Ti film is nitridated after opening simultaneously the Ga and N shutters to initiate the growth of GaN, while for samples B and C, a dedicated nitridation step is introduced before growth. The N flux used for substrate nitridation is the same one as for GaN growth. After the intentional substrate nitridaton process, the Ga shutter is opened to initiate the formation of GaN NWs. Further details concerning the nitridation process can be found in Ref. \citenum{Calabrese2019}. Table~\ref{tab:1} summarizes the GaN growth conditions employed for the growth of samples A--E. 

\subsection{Characterization} 
After growth, the NW morphology is investigated by field-emission scanning electron microscopy (SEM), XRD and cw-PL. The SEM analysis is carried out in a Zeiss Ultra 55 microscope using an acceleration voltage of 15~keV. The average NW radii for samples A--E, as determined from the analysis of several plan-view scanning electron micrographs using the open source software ImageJ,\citep{Schneider2012} are shown in Table~\ref{tab:1}.

Laboratory XRD measurements of $000L$ reflections are performed with Cu$K\alpha_{1}$ radiation (wavelength $\lambda=1.54056$~Å) using a Panalytical X-Pert Pro MRD$^{\mathrm{TM}}$ system equipped with a Ge~220 hybrid monochromator. To investigate the strain state of the NW ensembles, symmetric $\theta/2\theta$ scans are recorded across the GaN~$0002$, $0004$, and $0006$ Bragg reflections with a three-bounce Ge~$220$ analyzer crystal in front of the detector. The resolution functions for our experimental conditions are determined by measuring the same Bragg reflections of a bulk GaN$(0001)$ single crystal with a dislocation density lower than $10^{5}$~cm$^{-2}$ purchased from Ammono S.A.\citep{Kaganer2012}  For the determination of the NW tilt and twist, $\omega$ and $\varphi$ scans are recorded across the GaN~$0002$ and $1\bar{1}00$ reflections, respectively. These measurements are carried out without analyzer crystal, using a 1~mm slit in front of the detector.

GID measurements are performed at the beamline ID10 of the European Synchrotron Radiation Facility (ESRF) at an X-ray energy of 22~keV (wavelength $\lambda=0.5636$~Å). The grazing incidence angle is varied from $0.1^\circ$ to $0.6^\circ$. A linear detector (Mythen 1K, Dectris) is placed parallel to the substrate surface to cover the range of scattering angles $\psi$ around the diffraction angle of the GaN~$1\bar100$ reflection with $2\theta\approx 11.7^\circ$. The grazing exit angle is varied from $0.12^\circ$ to $0.33^\circ$. The sample is rotated about the substrate surface normal, and linear detector scans are recorded for different azimuthal angles $\omega$. The obtained reciprocal space maps in $\omega-\psi$ coordinates are similar to standard $\omega-2\theta$ maps measured in laboratory X-ray diffraction experiments.

Continuous-wave PL spectra are recorded utilizing a Horiba/Jobin-Yvon LabRam HR-800-UV spectrograph. The luminescence is excited at 9~K by the 325 nm line (photon energy 3.814 eV) of a Kimmon HeCd laser, dispersed in an 80~cm monochromator equipped with a 2400~lines/mm grating [resulting in a spectral resolution of 0.25~\AA\ (0.25 meV)], and detected by a liquid-N$_{2}$-cooled charge coupled device. The monochromator is calibrated by the spectral position of the atomic emission lines from Hg and Ne lamps. For the experiments presented here, the samples are measured side-by-side without moving the monochromator.

\begin{acknowledgement}
We thank Bernd Jenichen for his support with the analysis of the samples by XRD, Peter Zaumseil for his attempt of laboratory in-plane XRD measurements, Carsten Stemmler, Katrin Frank, and Michael H{ö}ricke for their dedicated maintenance of the MBE system, Anne-Kathrin Bluhm for her support with the scanning electron microscope, and Thomas Auzelle for a critical reading of the manuscript. Financial support provided by the Leibniz-Gemeinschaft under Grant SAW-2013-PDI-2 is gratefully acknowledged. P.\ C.\ acknowledges the funding from the Fonds National Suisse de la Recherche Scientifique through project 161032, and S.\ F.\ G.\ the partial financial support received through the Spanish program Ramón y Cajal (co-financed by the European Social Fund) under grant RYC-2016-19509 from Ministerio de Ciencia, Innovación y Universidades.
\end{acknowledgement}

\bibliography{acsn_bib}

\end{document}